\documentclass[oneside,onecolumn]{elsart}
\usepackage{amsfonts}
\usepackage{amsmath}
\usepackage{graphicx}

\input{tcilatex}

\begin{document}

\begin{frontmatter}

\title{Geometric measures of redundance and irrelevance tradeoff exponent to 
choose suitable delay times for continuous systems}

\author{Xiaodong Luo\thanksref{correspond}} 
\author{Michael Small}

\address{Department of Electronic and Information Engineering, Hong Kong 
Polytechnic University, Hung Hom, Hong Kong.}
\thanks[correspond]{Corresponding author. Tel: +852 2766 6199,
Fax: +852 2362 8439, email: {\tt enxdluo@eie.polyu.edu.hk}.}

\date{\today}

\begin{abstract}
  Using the concept of the geometric measures of \textit{redundance and irrelevance% 
tradeoff exponent (RITE)}, we present a new method to determine suitable 
delay times for continuous systems. After applying the \textit{RITE} algorithm to both 
simulation and experimental observations, we find the results obtained are close 
to those obtained from the criterion of the \textit{average mutual information (AMI)}, 
while the \textit{RITE} algorithm has the following advantages: simple implementation, 
reasonable computational cost and robust performance against observational noise.
\end{abstract}

\end{frontmatter}

\section{INTRODUCTION}

Since the embedding theorem of Takens \cite{takens} appeared, a number of
papers have been published on criteria for estimating a suitable delay time
for a nonlinear time series. One criterion, based on the \textit{second
order autocorrelation }(\textit{SOAC}), chooses the time as the delay when
the \textit{SOAC} first becomes zero or drops to a certain fraction of its
initial value \cite{Albano 1}. This method is simple to implement, but it
lacks a universal fraction for different systems to obtain suitable delay
times \footnote{%
For example, the first zero criterion \ is successful for the R\"{o}ssler
system but it fails for the Lorenz system.}. As a generalisation of the
above idea, Albano \textit{et al. }\cite{Albano2} \ proposed a heuristic
idea. They take the time at the consistent extrema of different higher-order
autocorrelation functions as candidates for a suitable embedding window,
therefore if we choose an embedding dimension, we also choose a delay time.
As a further step, having noticed that the \textit{SOAC} is actually a
linear measure of dependence, Fraser \& Swinney \cite{Fraser} introduced an
important statistic, \textit{mutual information}, based on information
theory. \textit{Mutual information} is a nonlinear measure of dependence
between two data sets, for a scalar time series, we can use the \textit{%
average mutual information} \textit{(AMI) }to select a proper delay time.
The criterion is to take the time at the first local minimum of the \textit{%
AMI} as the desired delay time. \textit{Mutual information} is a valuable
concept, but it is rather complex to implement. In addition, it was found
its performance was not very robust for small data sets \cite{Martinerie}.

From other viewpoints, some criteria were proposed based on the utilization
of the geometric information of the reconstructed attractor in embedding
space. Buzug and Pfister \cite{Buzug} devised the \textit{fill-factor
algorithm} to determine a suitable delay time by examining the attractor's
expansion in embedding space. It will be selected as the suitable delay time
when the fill-factor is maximized. But this algorithm also takes into
account the situation of \textquotedblright overfolding\textquotedblright ,
and more seriously, it will fail to yield significant delays if the
attractor has more than one unstable focus \cite{Buzug}. As a solution,
Buzug and Pfister designed another algorithm, \textit{integral local
deformation (ILD)}. This algorithm will choose a suitable delay time when
the attractor's local minimum deformation is achieved . Comparatively, this
algorithm needs substantially more computational time than the \textit{%
fill-factor} algorithm, and as we will indicate in the later section, it
might be more suitable to use this algorithm to choose embedding window
rather than delay time.

There have been many other criteria proposed. For example, Rosenstein \cite%
{Rosenstein} developed an approach named \textit{reconstruction signal
strength }resting on the concepts of redundance error and irrelevance error%
\textit{, }their approach is computationally efficient and can obtain a
satisfactory performance, but the criterion for choosing suitable delay
times is somewhat empirical. In this communication we do not intend to
provide a detailed review, readers are invited to refer to the literature 
\cite{Liebert1} and \cite{Kember} and references therein for more details.

In the remaining sections, firstly we will propose a new algorithm to choose
suitable delay times based on the concept of the geometric measures of 
\textit{redundance and irrelevance tradeoff exponent} (\textit{RITE}). Then
we will examine the performance of this algorithm by applying it to data
sets from both simulation and experimental observations. Finally we have a
summary.

\section{THE ALGORITHM OF RITE}

In a very recent paper Cellucci and coworkers \cite{cellucci and albano
comparative} state their viewpoint on embedding methods as: \textit{A
circular logic has resulted in which embedding criteria are assessed by an
adjudicating criterion which is itself an embedding criterion}. Following
this viewpoint, we learn that the best embedding criteria might differ under
different adjudicating criteria. Hence we would like to elucidate that we\
do not seek the best embedding criteria for different adjudicating criteria,
instead we intend to let our adjudicating criterion fit as many cases as
possible. \ 

As we have known, sufficiently high embedding dimension is a necessary but
not sufficient condition to form an embedding reconstruction according to
the embedding theorem of Takens. To be an embedding by itself will impose
two constraints on the reconstruction mapping $\Psi :$ $%
%TCIMACRO{\U{211d} }%
%BeginExpansion
\mathbb{R}
%EndExpansion
\rightarrow 
%TCIMACRO{\U{211d} }%
%BeginExpansion
\mathbb{R}
%EndExpansion
^{m}$, where $m$ is embedding dimension. One is that $\Psi $ shall be a
one-to-one mapping, the other is that the \textit{derivative mapping} $%
D\cdot \Psi $ shall also be one-to-one \cite{Galka}, where $D$ denotes the
differentiating operator on $\Psi $.

In practice, although some delay times no longer lead to an embedding
reconstruction (unlike the ideal situations), which will be discussed in the
following content, it is hoped there are at least some others remaining. We
note that, these remaining delay times equivalently lead to an embedding in
the sense of characterizing the reconstructed attractor, although some
particular values might indeed facilitate the analysis of a time series.
Hence our adjudicating criterion is to guarantee the reconstruction mapping
to be an embedding, and even if we obtain different delay times from
different algorithms, we still consider them as suitable candidates for an
embedding reconstruction.

\begin{figure*}[t]
\centering
\includegraphics[width=5.5in]{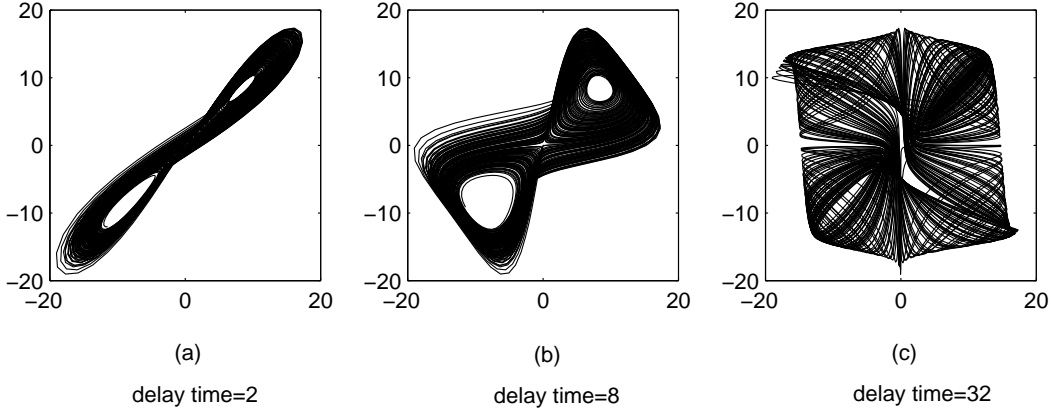}
\caption{Effects of different delay times on the reconstructed attractor of the Lorenz system
in the two-dimensional embedding space. (a) delay time=2; (b) delay time=8; (c) delay time= 32.}
\label{delaytime}
\end{figure*}

For a delay time embedding reconstruction, a scalar time series $%
\{x_{i}:i=1,2,\ldots ,N\}$ is used to construct vectors $\overrightarrow{%
X_{i}}=(x_{i},x_{i+\tau },\ldots ,x_{i+(m-1)\tau })$ in $%
%TCIMACRO{\U{211d} }%
%BeginExpansion
\mathbb{R}
%EndExpansion
^{m}$, where $m$ is embedding dimension and $\tau $ is delay time. Now let
us consider the effects of different delay times on the reconstructed
attractor. Without losing generality, we confine our discussions to the
two-dimensional embedding space $x_{i+\tau }$ vs. $x_{i}$. Fig. \ref%
{delaytime} demonstrates the reconstructed attractors of the Lorenz system 
\cite{Galka} for three different delay times. When $\tau $ is too small,
then $x_{i+\tau }$ will be very close to $x_{i}$ due to the continuity of
the manifold. Therefore the pair points $\left( x_{i},x_{i+\tau }\right) $
will distribute around the unity line $x_{i+\tau }=x_{i}$ as indicated in
Fig. \ref{delaytime} (a). But in practice, the presence of noise will let an
embedding vector $\overrightarrow{X}_{i}=\left( x_{i},x_{i+\tau },\ldots
,x_{i+(m-1)\tau }\right) $ distributed as a "ball" rather than a point in
metric space $%
%TCIMACRO{\U{211d} }%
%BeginExpansion
\mathbb{R}
%EndExpansion
^{m}$. The balls of adjacent vectors might intersect with each other, hence
the reconstruction mapping $\Psi :$ $%
%TCIMACRO{\U{211d} }%
%BeginExpansion
\mathbb{R}
%EndExpansion
\rightarrow 
%TCIMACRO{\U{211d} }%
%BeginExpansion
\mathbb{R}
%EndExpansion
^{m}$ is not one-to-one and no longer an embedding. When delay time $\tau $
is too large, say $\tau =32$ as adopted in Fig. \ref{delaytime} (c), the
reconstructed attractor is overfolded and does not preserve the geometric
structure of the original attractor comparing to those in panel (a) and (b),
which also means the reconstruction is not an embedding.%
%TCIMACRO{%
%\TeXButton{delaytime}{\begin{figure*}[t]
%\centering
%\includegraphics[width=5.5in]{delaytime.eps}
%\caption{Effects of different delay times on the reconstructed attractor of the Lorenz system
%in the two-dimensional embedding space. (a) delay time=2; (b) delay time=8; (c) delay time= 32.}
%\label{delaytime}
%\end{figure*}}}%
%BeginExpansion
%EndExpansion

From the viewpoint of information theory, delay time $\tau $ is too small
means that $x_{i+\tau }$ contain mainly redundant information of $x_{i}$.
This is called \textit{redundance}. If delay time is too large, then for
chaotic systems, $x_{i+\tau }$ will be irrelevant to $x_{i}$, hence $%
x_{i+\tau }$ contains no information of $x_{i}$. This is known as \textit{%
irrelevance}.\textit{\ }As\textit{\ }Liebert and Schuster \cite{Liebert2}
have argued that, we shall consider not only the effect of redundance but
also that of irrelevance in estimate of suitable delay times. Therefore a
tradeoff shall be achieved between redundance and irrelevance so as to
guarantee the reconstruction mapping to be an embedding. We define following
statistic, namely \textit{redundance and irrelevance tradeoff exponent (RITE)%
}, to measure the tradeoff, 
\begin{equation}
RITE=\frac{\rho (x_{i},x_{i+\tau })\left\langle {\small x}%
_{i}^{2}\right\rangle +(1-\rho (x_{i},x_{i+\tau }))\left\langle {\small x}%
_{i}\right\rangle ^{2}}{\left\langle {\small x}_{i}^{2}\right\rangle
+\left\langle {\small x}_{i}\right\rangle ^{2}}  \label{rite}
\end{equation}%
where $\left\langle \cdot \right\rangle $ denotes the expectation taken over
time $i$ and 
\begin{equation}
\rho (x_{i},x_{i+\tau })=\frac{cov(x_{i},x_{i+\tau })}{var(x_{i})}=\frac{%
\left\langle x_{i}x_{i+\tau }\right\rangle -\left\langle x_{i}\right\rangle
^{2}}{\left\langle x_{i}^{2}\right\rangle -\left\langle x_{i}\right\rangle
^{2}}  \label{autocorrelation}
\end{equation}%
where $\rho (x_{i},x_{i+\tau })$ is the \textit{SOAC}, $cov(x_{i},x_{i+\tau
})$ and $var(x_{i})$ are the covariance with delay time $\tau $ and the
variance of the time series $\{x_{i}\}$ respectively. After simplifications,
we have 
\begin{equation}
RITE=\frac{\left\langle x_{i}x_{i+\tau }\right\rangle }{\left\langle
x_{i}^{2}\right\rangle +\left\langle x_{i}\right\rangle ^{2}}
\label{simplification}
\end{equation}

As we shall see in the following content, Eqn. (\ref{simplification}) is
only a constant affine transformation of the \textit{SOAC }if directly
applied to the original time series $\{x_{i}\}$. Before that let us first
interpret the meaning of Eqn. (\ref{rite}). We take $\left\langle
x_{i}^{2}\right\rangle $ as the case of complete redundance for the measure $%
\left\langle x_{i}x_{i+\tau }\right\rangle $, when delay time $\tau $ tends
to zero and referring to $x_{i+\tau }$ brings no more information of $x_{i}$%
. Conversely, $\left\langle x_{i}\right\rangle ^{2}$ is the case of complete
irrelevance for the measure $\left\langle x_{i}x_{i+\tau }\right\rangle $,
when $x_{i+\tau }$ is irrelevant and thus uncorrelated to $x_{i}$, hence $%
\left\langle x_{i}x_{i+\tau }\right\rangle $ is reduced to $\left\langle
x_{i}\right\rangle ^{2}$. The \textit{SOAC} $\rho (x_{i},x_{i+\tau })$ plays
the role to measure the redundance between $x_{i+\tau }$ and $x_{i}$ with a
weight of $\left\langle {\small x}_{i}^{2}\right\rangle \left/ \left\langle 
{\small x}_{i}^{2}\right\rangle +\left\langle {\small x}_{i}\right\rangle
^{2}\right. $, while $1-\rho (x_{i},x_{i+\tau })$ denotes the measure of
irrelevance with the assigned weight of $\left\langle x_{i}\right\rangle
^{2}\left/ \left\langle {\small x}_{i}^{2}\right\rangle +\left\langle 
{\small x}_{i}\right\rangle ^{2}\right. $. Starting from $\tau =0$, as delay
time $\tau $ increases, the redundance measure $\rho (x_{i},x_{i+\tau })$
shall usually decrease while the irrelevance measure $1-\rho
(x_{i},x_{i+\tau })$ shall increase, hence a natural criterion is to choose
the suitable delay time at the first local minimum of \textit{RITE}, which
guarantees the reconstruction to be an embedding in an optimal way according
to Eqn. (\ref{rite}).

If directly applying the measure of \textit{RITE }to measure the original
scalar time series $\{x_{i}\}$, we can find from Eqn. (\ref{rite}) it is a
trivial measure with the same performance as that of the \textit{SOAC }since 
$\left\langle x_{i}^{2}\right\rangle $ and $\left\langle x_{i}\right\rangle
^{2}$ are both independent of delay time $\tau $. A remedy is that, we can
equivalently characterize the reconstructed attractor in the two-dimensional
embedding space $x_{i+\tau }$ vs. $x_{i}$ instead of in the time domain.

\begin{figure*}[t]
\centering
\includegraphics[width=4in]{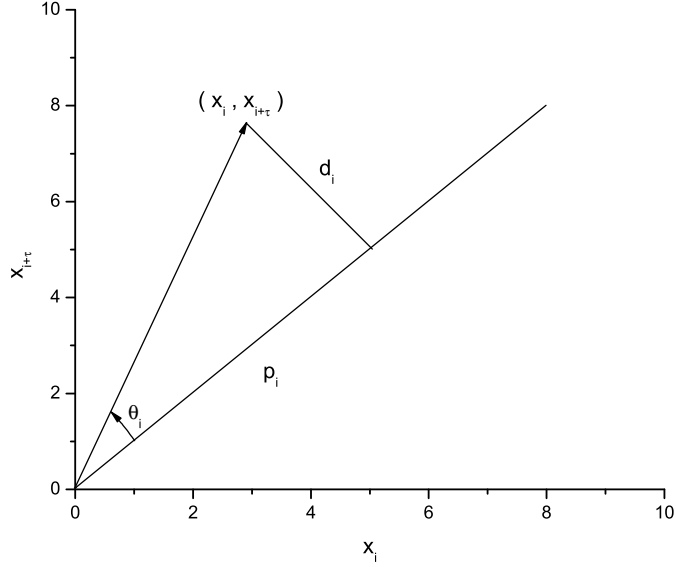}
\caption{Geometric variables in the two-dimensional embedding space.}
\label{line}
\end{figure*}

Let $\overrightarrow{(x_{i,}x_{i+\tau })}$ denote the vector from the origin
to point $(x_{i},x_{i+\tau })$ in the two dimensional embedding space, as
shown in Fig. \ref{line}, we have the distance $d_{i}$ of the pair points $%
(x_{i},x_{i+\tau })$ to the identity line $x_{i+\tau }=x_{i}$ expressed by:%
%TCIMACRO{%
%\TeXButton{line}{\begin{figure*}[t]
%\centering
%\includegraphics[width=4in]{line.eps}
%\caption{Geometric variables in the two-dimensional embedding space.}
%\label{line}
%\end{figure*}} }%
%BeginExpansion
%EndExpansion
\begin{equation}
d_{i}=\frac{1}{\sqrt{2}}\left\vert x_{i+\tau }-x_{i}\right\vert 
\label{distance}
\end{equation}%
where $\left\vert \cdot \right\vert $ denotes the distance in Euclidean
space. The projection length $p_{i}$ of vector $\overrightarrow{%
(x_{i,}x_{i+\tau })}$ onto the identity line is: 
\begin{equation}
p_{i}=\frac{1}{\sqrt{2}}\left\vert x_{i+\tau }+x_{i}\right\vert 
\label{projection}
\end{equation}%
Therefore the angle between vector $\overrightarrow{(x_{i,}x_{i+\tau })}$
and the identity line is : 
\begin{equation}
\theta _{i}=\tan ^{-1}\left\vert \frac{x_{i+\tau }-x_{i}}{x_{i+\tau }+x_{i}}%
\right\vert   \label{angle}
\end{equation}

From Eqn. (\ref{distance}), (\ref{projection}) and (\ref{angle}), we obtain
three new time series $\{d_{i}\}$, $\{p_{i}\}$ and $\{\theta _{i}\}$ derived
from the original one which consist of geometric description variables of
the reconstructed attractor in the two-dimensional embedding space. These
geometric variables shall also be continuous in the time domain since all of
the three above transforms are continuous. We apply the measure of \textit{%
RITE }to these geometric variables with the same criterion to choose
suitable delay times as having stated above, i.e., a suitable delay time
will be chosen at the first local minimum of the geometric measures of 
\textit{RITE}.

\section{NUMERICAL RESULTS}

%TCIMACRO{%
%\TeXButton{lorenzDeformation}{\begin{figure*}[t]
%\centering
%\includegraphics[width=2.5in]{lorenzDeformation.eps}
%\includegraphics[width=2.5in]{shortLorenzDeformation.eps}
%\caption{Figure in the left panel indicates the average integral local deformation
%vs. delay time for  the time series from the Lorenz system with 9000 data points. 
%The number of reference points is 500, radius for neighbour searching is set to 5. 
%Embedding dimension $m$ varies from 2 to 6 (from upper to lower) and delay time 
%increases from 1 to 50. Figure in the right panel adopts the same parameters as the left for
%calculations except that the time series is shorter, contisting of only 1200 data points.}
%\label{lorenzDeformation}
%\end{figure*}}}%
%BeginExpansion
\begin{figure*}[t]
\centering
\includegraphics[width=2.5in]{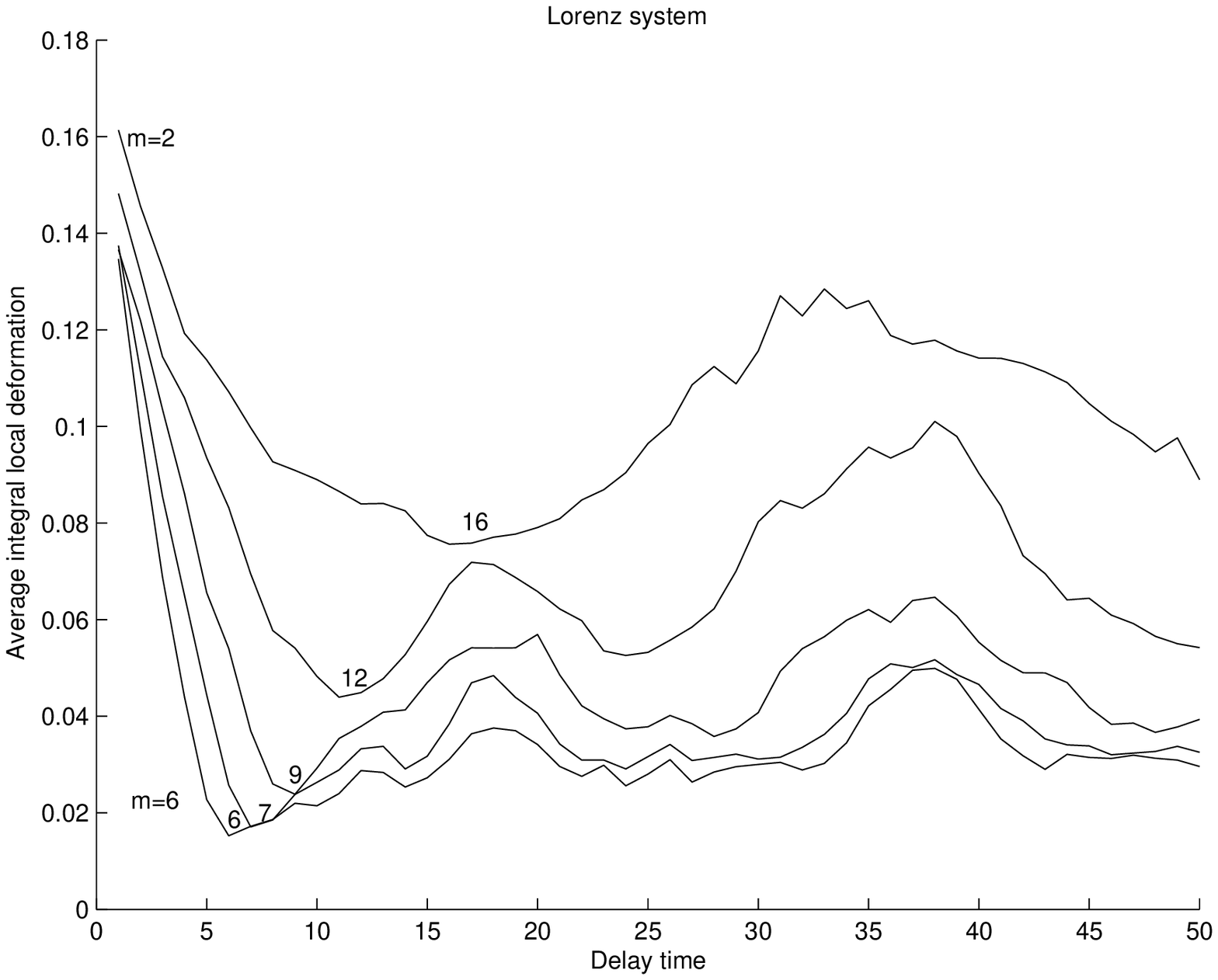}
\includegraphics[width=2.5in]{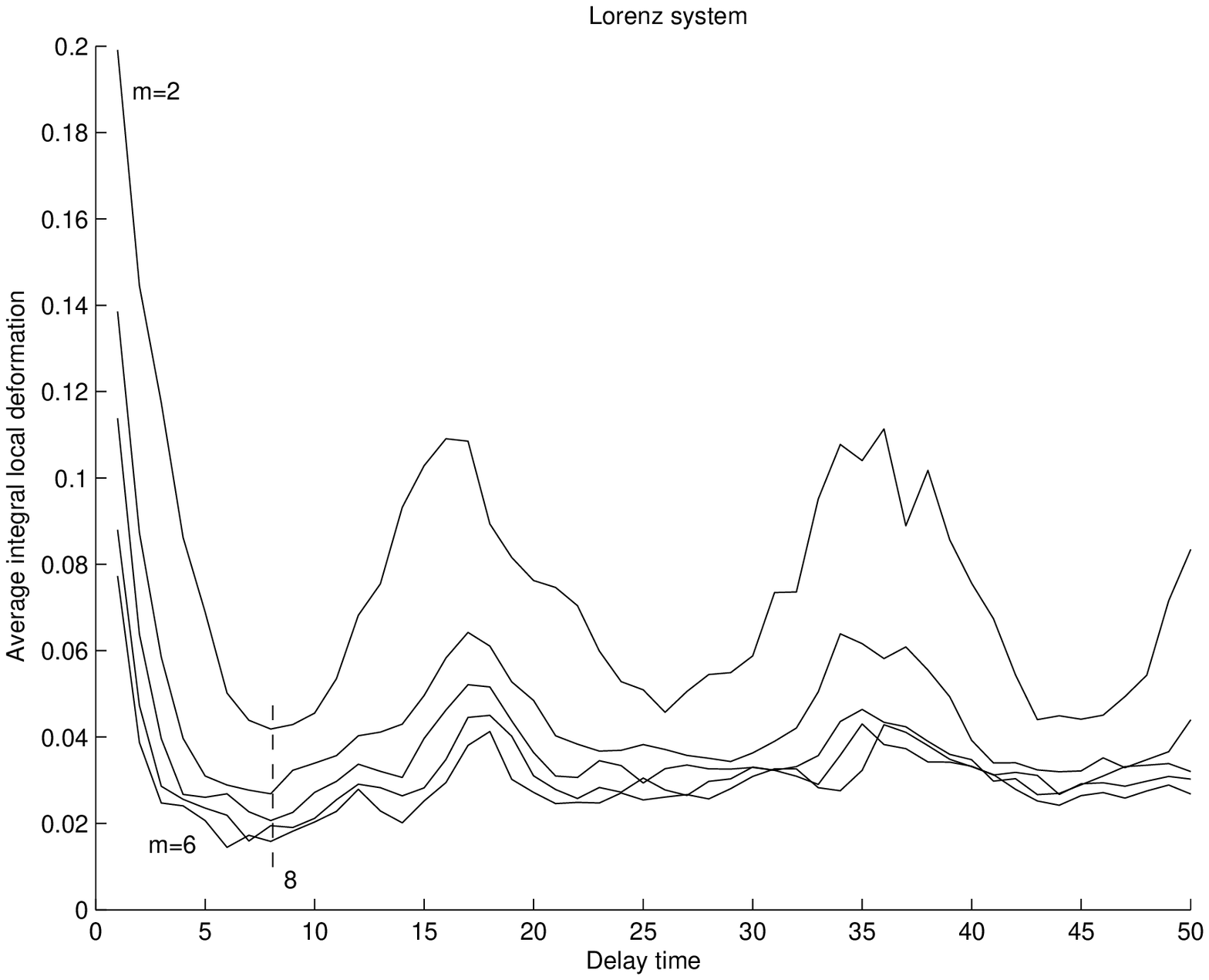}
\caption{Figure in the left panel indicates the average integral local deformation
vs. delay time for  the time series from the Lorenz system with 9000 data points. 
The number of reference points is 500, radius for neighbour searching is set to 5. 
Embedding dimension $m$ varies from 2 to 6 (from upper to lower) and delay time 
increases from 1 to 50. Figure in the right panel adopts the same parameters as the left for
calculations except that the time series is shorter, contisting of only 1200 data points.}
\label{lorenzDeformation}
\end{figure*}%
%EndExpansion

We note that if the origin in the embedding space of a time series is
marginal to or even outside of the reconstructed attractor, the sensitivity
of the geometric measures of RITE to different delay times will be
significantly reduced. We therefore conduct the following smooth affine
transform on the original time series \{$x_{i}$\}. 
\begin{equation}
y_{i}=\frac{x_{i}-\left\langle x_{i}\right\rangle }{\sqrt{var(x_{i})}}
\label{transformation}
\end{equation}

The new time series \{$y_{i}$\} shall have the same dynamical properties in
the time domain as the original time series has, while it takes the origin
of embedding space as the \textquotedblright center\textquotedblright\ of
its reconstructed attractor in a statistical sense. With this consideration,
we prefer to studying the time series \{$y_{i}$\} rather than \{$x_{i}$\}.
In addition, we will discard the scale factor $1\left/ \sqrt{2}\right. $ of
both Eqn. (\ref{distance}) and (\ref{projection}) in all of our calculations
without affecting the results.

We will study the simulation data sets from the Lorenz and R\"{o}ssler
systems \cite{Galka}. For the Lorenz system, the equations are: 
\begin{equation}
\left\{ 
\begin{array}{l}
\dot{x}(t)=\sigma (y(t)-x(t)) \\ 
\dot{y}(t)=rx(t)-y(t)-x(t)z(t) \\ 
\dot{z}(t)=x(t)y(t)-bz(t)%
\end{array}%
\right.  \label{lorenz}
\end{equation}%
with parameters $\sigma =10$, $r=28$, $c=8\left/ 3\right. $ and the sampling
time $\Delta t_{s}=0.02s$. For the R\"{o}ssler system, the equations are: 
\begin{equation}
\left\{ 
\begin{array}{l}
\dot{x}(t)=-y(t)-z(t) \\ 
\dot{y}(t)=x(t)+ay(t) \\ 
\dot{z}(t)=b+z(t)(x(t)-c)%
\end{array}%
\right.  \label{rossler}
\end{equation}%
with parameters $a=0.15$, $b=0.20$, $c=10.00$ and the sampling time $\Delta
t_{s}=0.1s$.\ 
%TCIMACRO{%
%\TeXButton{table1}{\begin{table}[h]
%\centering
%\parbox{4in}{\caption{\label{table1} Delay times chosen by the algorithms of \textit{ILD}, \textit{
%AMI} and the geometric measures of \textit{RITE}.}}
%\begin{tabular*}{4in}{cccccc}
%\hline\hline
%Data set & \textit{ILD} & \textit{AMI} & \multicolumn{3}{c}{Geometric
%measures of \textit{RITE}} \\ \cline{4-6}
%& $\tau /m$ &  & distance & projection & angle \\ \hline
%\multicolumn{1}{l}{Lorenz} & 9/4 & 8 & 9 & 12 & 10 \\ 
%\multicolumn{1}{l}{R\"{o}ssler} & 10/3 & 16 & 16 & 14 & 13 \\ 
%\multicolumn{1}{l}{Sunspot} & 2 & 3 & 2 & 3 & 2 \\ 
%\multicolumn{1}{l}{S4} & \multicolumn{1}{c}{5/5} & \multicolumn{1}{c}{8} & 
%\multicolumn{1}{c}{7} & \multicolumn{1}{c}{8} & \multicolumn{1}{c}{5} \\ 
%\hline
%\end{tabular*}\end{table}}}%
%BeginExpansion
\begin{table}[h]
\centering
\parbox{4in}{\caption{\label{table1} Delay times chosen by the algorithms of \textit{ILD}, \textit{
AMI} and the geometric measures of \textit{RITE}.}}
\begin{tabular*}{4in}{cccccc}
\hline\hline
Data set & \textit{ILD} & \textit{AMI} & \multicolumn{3}{c}{Geometric
measures of \textit{RITE}} \\ \cline{4-6}
& $\tau /m$ &  & distance & projection & angle \\ \hline
\multicolumn{1}{l}{Lorenz} & 9/4 & 8 & 9 & 12 & 10 \\ 
\multicolumn{1}{l}{R\"{o}ssler} & 10/3 & 16 & 16 & 14 & 13 \\ 
\multicolumn{1}{l}{Sunspot} & 2 & 3 & 2 & 3 & 2 \\ 
\multicolumn{1}{l}{S4} & \multicolumn{1}{c}{5/5} & \multicolumn{1}{c}{8} & 
\multicolumn{1}{c}{7} & \multicolumn{1}{c}{8} & \multicolumn{1}{c}{5} \\ 
\hline
\end{tabular*}\end{table}%
%EndExpansion
\ \ 

We will also apply the geometric measures of \textit{RITE} to the sunspot
record from year 1700 to year 1987 and infant respiratory data during stage $%
4$ sleep (S4) \cite{Michael}. In addition, we will calculate delay times
chosen by the \textit{ILD} and \textit{AMI} algorithms for the comparison
purpose. Our results are listed in Table \ref{table1}.\ $\ \ \ $

Although the \textit{ILD} algorithm was originally designed to determine
suitable delay times $\tau $, it might be more appropriate to utilize it in
establishing embedding window $m\cdot \tau $. As indicated in the left panel
of Fig. \ref{lorenzDeformation}, when using Eqn. (24) in Ref. \cite{Buzug}
for calculations, for the Lorenz system the products of embedding dimensions 
$m$ ($m>$correlation dimension $d_{c}$) and their corresponding delay times $%
\tau $ at the first local minimum of the average \textit{ILD }are nearly a
constant of 36. This conclusion also holds for data sets of the R\"{o}ssler
system and S4. In contrast, the sunspot record has consistent local minima
and the products of $m\cdot \tau $ do not keep constant. This still does not
contradict our conclusion as the sunspot record is an extremely short time
series. As shown in the right panel of Fig. \ref{lorenzDeformation}, the
constant embedding window will vanish when the time series from the Lorenz
system is shorter, instead a consistent local minimum appears at $\tau =8$. 
%TCIMACRO{%
%\TeXButton{nlpe}{\begin{figure*}[t]
%\centering
%\includegraphics[width=4in]{NLPE.eps}
%\caption{Nonlinear prediction error of local constant model v.s. delay time. Embedding 
%dimensions used in the model are 4,3,3 and 5 for the Lorenz system, the R\"{o}ssler 
%system, the sunspot record and data set of S4 respectively. The ranges of delay time are 
%all from 1 to 60 . The \textit{NLPE}s corresponding to delay times in Table \ref{table1} chosen by 
%different algorithms for each data set are marked with diamonds. We use the program 
%\textit{zeroth} in TISEAN package \cite{Tisean} for our calculations.}
%\label{nlpe}
%\end{figure*}}}%
%BeginExpansion
\begin{figure*}[t]
\centering
\includegraphics[width=4in]{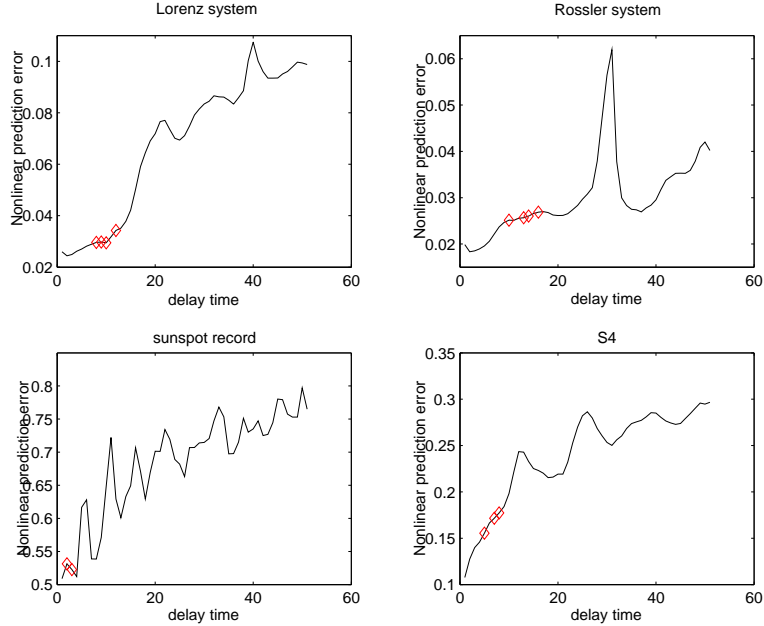}
\caption{Nonlinear prediction error of local constant model v.s. delay time. Embedding 
dimensions used in the model are 4,3,3 and 5 for the Lorenz system, the R\"{o}ssler 
system, the sunspot record and data set of S4 respectively. The ranges of delay time are 
all from 1 to 60 . The \textit{NLPE}s corresponding to delay times in Table \ref{table1} chosen by 
different algorithms for each data set are marked with diamonds. We use the program 
\textit{zeroth} in TISEAN package \cite{Tisean} for our calculations.}
\label{nlpe}
\end{figure*}%
%EndExpansion

Since embedding window $m\cdot \tau $ remains constant, different delay
times will be obtained from the \textit{ILD} algorithm for different
embedding dimensions, nevertheless, we think the \textit{ILD} algorithm can
still indicate how to obtain a proper embedding reconstruction with
sufficiently high embedding dimension. Firstly we need to choose an optimal
embedding dimension for each data set (except for the sunspot record) under
the criterion of\textit{\ global} \textit{False Nearest Neighbours (GFNN)} 
\cite{Kennel} \cite{cellucci and albano comparative}, then we can obtain the
corresponding delay time according to embedding window \footnote{%
It has to admit it is somewhat \textquotedblright
circular\textquotedblright\ in this situation,since the choice of the
optimal embedding dimension by \textit{GFNN} algorithm in turn needs to take
the suitable delay time as a parameter . In our calculation, we use the
suitable delay time obtained by the \textit{AMI} algorithm in Table \ref%
{table1} as the parameter to determine the optimal embedding dimension for
each data set (except for the sunspot record).}. For the sunspot record we
choose delay time at the first consistent local minimum of the average 
\textit{ILD}. The results are indicated in Table \ref{table1}.

From Table \ref{table1} we find that, loosely speaking, the results of
different algorithms are close to each other. As we have stated in the
previous section, although delay times chosen by different geometric
measures of \textit{RITE} and the other two algorithms are usually
different, we still take all of them as the suitable candidates for an
embedding reconstruction.

We use the \textit{nonlinear prediction error (NLPE)} to verify the
reconstruction quality of our choice. As we have known, local constant model 
\cite{Farmer} utilises nearest neighbours for nonlinear prediction, when
sufficiently high embedding dimension is reached, most of the effect of
false nearest neighbours will be excluded. When embedding dimension and the
radius of neighbour searching are fixed, the \textit{NLPE} will only depend
on delay time. Hence the \textit{NLPE} can qualitatively determine whether
our choice for an embedding reconstruction is acceptable, as the prediction
error of a suitable delay time shall achieve a tradeoff between being too
small and being too large if the time series is not completely predictable
or completely unpredictable. In Fig. \ref{nlpe}, the \textit{NLPE}s
corresponding to delay times listed in Table \ref{table1} chosen by
different algorithms for each data set are marked with diamonds. As we can
find, certain tradeoff for each geometric measure of \textit{RITE} is indeed
achieved.

Now let us examine the computational cost of each algorithm listed in Table
1. Let $N$ denotes the data set size of time series $\{x_{i}\}$, then the 
\textit{ILD} algorithm approximately requires $O(N_{ref}\times (N\ln N))$
unit operations on searching nearest neighbours for each embedding dimension
and each delay time, where $N_{ref}$ is the number of reference points. The 
\textit{AMI} algorithm needs about $O(N^{2})$ unit operations to calculate
joint probability distribution for each delay time, while the \textit{RITE}
algorithm will be faster than both of them, undergoing about $O(N)$ unit
operations on both the transforms over the original data set and the
calculations of expectation for each delay time. \ 
%TCIMACRO{%
%\TeXButton{table2}{\begin{table*}[t]
%\centering
%\parbox{5.5in}{\caption{\label{table2} Delay times chosen by the geometric measures of \textit{RITE} for the time 
%series from the Lorenz and R\"{o}ssler systems contaminited with observational Gaussian white noises .}}
%\begin{tabular*}{5.5in}{ccccccc}
%\hline\hline
%Noise level (\%) & \multicolumn{3}{c}{Lorenz system} & \multicolumn{3}{c}{R\"{o}ssler system} \\ 
%& distance & projection & angle & distance & projection & angle \\ \hline
%0 & 9 & 12 & 10 & 16 & 14 & 13 \\ 
%3 & 9 & 12 & 10 & 16 & 14 & 13 \\ 
%6 & 9 & 12 & 9 & 16 & 14 & 13 \\ 
%9 & 9 & 12 & 10 & 16 & 14 & 13 \\ 
%12 & 9 & 12 & 8 & 16 & 14 & 2 \\ \hline
%\end{tabular*}\end{table*}}}%
%BeginExpansion
\begin{table*}[t]
\centering
\parbox{5.5in}{\caption{\label{table2} Delay times chosen by the geometric measures of \textit{RITE} for the time 
series from the Lorenz and R\"{o}ssler systems contaminited with observational Gaussian white noises .}}
\begin{tabular*}{5.5in}{ccccccc}
\hline\hline
Noise level (\%) & \multicolumn{3}{c}{Lorenz system} & \multicolumn{3}{c}{R\"{o}ssler system} \\ 
& distance & projection & angle & distance & projection & angle \\ \hline
0 & 9 & 12 & 10 & 16 & 14 & 13 \\ 
3 & 9 & 12 & 10 & 16 & 14 & 13 \\ 
6 & 9 & 12 & 9 & 16 & 14 & 13 \\ 
9 & 9 & 12 & 10 & 16 & 14 & 13 \\ 
12 & 9 & 12 & 8 & 16 & 14 & 2 \\ \hline
\end{tabular*}\end{table*}%
%EndExpansion

We will also test the robustness of the geometric measures of \textit{RITE}
against observational Gaussian white noise $N(0,\delta ^{2})$. Noise level
is defined as the ratio of $\delta $ to $\delta _{s}$, where $\delta _{s}$
is the standard deviation of the original scalar time series $\{x_{i}\}$
before the transform of Eqn. (\ref{transformation}). As indicated in Table %
\ref{table2}, using delay times chosen at noise level zero as the
references, we find both the distance and the projection measures of \textit{%
RITE }are rather robust against observational noise, noise level up to 12\%
still can not affect the choices of delay time. As expected, the angle
measure of \textit{RITE }is more sensitive to noise. For the Lorenz system,
small fluctuations of the choice appear when noise level is higher than 6\%.
For the R\"{o}ssler system, the performance seems better. The odd choice $%
\tau =2$ at noise level 12\% follows our criterion suggested above, which is
due to a small spike on the curve of the angle measure of \textit{RITE} vs.
delay time, while the next local minimum is exactly at delay time $\tau =13$%
. Although the robustness against observational noise of the geometric
measures of \textit{RITE} might vary from system to system, we believe in
general it is satisfactory.

\section{CONCLUSION}

It has been a difficult problem to set up a universal criterion for the
choice of delay time. \textit{Average} \textit{mutual information} is the
most preferred statistic used for choosing delay time since it has a
valuable physical meaning, but it requires a complicated implementation
algorithm. To achieve higher accuracy, more complex implementation and more
running time are needed. Also it does not deal very well with short time
series. Comparatively, the \textit{RITE} algorithm intends to provide an
optimal choice of delay time with the objective to guarantee the
reconstruction to be an embedding. Our calculates indicate that the \textit{%
RITE }algorithm performs well on a variety of time series of various lengths
and even in the presence of substantial noise. We therefore feel that such a
simple algorithm should be preferred to the more complex implementation
suggested previously.

\section{ACKNOWLEDEGMENT}

This research was supported by a Hong Kong Polytechnic University Research
Grant (No. A-PE46).


\begin{thebibliography}{10}
\bibitem[1]{takens} F. Takens, in D. A. Rand, and L. S. Young, editors, 
\textit{Dynamical Systems and Turbulence}, Lecture Notes in Mathematics Vol.
898 (Springer-Verlag, New York, 1980).

\bibitem[2]{Albano 1} A. M. Albano, J. Muench, C. Schwartz, A. I. Mees, and
P. E. Rapp. Singular value decomposition and the Grassberger-Procaccia
algorithm. \textit{Phys. Rev. A} 38, 3017 (1988).

\bibitem[3]{Albano2} A. M. Albano, A. Passamante, and M. E. Farrell, Using
higher-order correlations to define an embedding window,\textit{\ Physica D}
54, 85 (1991).\textit{\ }

\bibitem[4]{Rosenstein} M. T. Rosenstein, J. J. Collins, and C. J. De Luca,
Reconstruction expansion as a geometry-based framework for choosing proper
delay times\textit{, Physica D} 73, 82 (1994).

\bibitem[5]{Fraser} A. M. Fraser, and H. L. Swinney, Independent coordinates
for strange attractors from mutual information, \textit{Phys. Rev. A} 33,
1134 (1986).

\bibitem[6]{Galka} A. Galka, \textit{Topics in Nonlinear Time Series
Analysis with Implications for EEG Analysis, Advanced Series in Nonlinear
Dynamics}, Vol. 14 (World Scientific, 2000).

\bibitem[7]{Martinerie} J. M. Martinerie, A. M. Albano, A. I. Mees, and R.
E. Rapp, Mutual information, strange attractors, and the optimal estimation
of dimension, \textit{Phys. Rev. A} 45, 7058 (1992).

\bibitem[8]{Buzug} T. Buzug, and G. Pfister, Optimal delay time and
embedding dimension for delay-time coordinates by analysis of the global
static and local dynamical behavior of strange attractors, \textit{Phys.
Rev. A} 45, 7073 (1992).

\bibitem[9]{Liebert1} W. Liebert, K. Pawelzik, and H. G. Schuster, Optimal
embeddings of chaotic attractors from topological considerations, \textit{%
Europhys. Lett.} 14, 521 (1991).

\bibitem[10]{Kember} G. Kember, and A. C. Fowler, A correlation function for
choosing time delays in phase portrait reconstructions, \textit{Phys. Lett. A%
} 179, 72 (1993).

\bibitem[11]{cellucci and albano comparative} C. J. Cellucci, A. M. Albano,
and P. E. Rapp, Comparative study of embedding methods, \textit{Phys. Rev. E}
67, 066210 (2003).

\bibitem[12]{Liebert2} W. Liebert, and H. Schuster, Proper choice of the
time delay for the analysis of chaotic time series, \textit{Phys. Lett. A }%
143, 107 (1989).

\bibitem[13]{Michael} M. Small, K. Judd, M. Lowe, and S. Stick. Is breathing
in infants chaotic? Dimension estimates for respiratory patterns during
quiet sleep, \textit{Journal of Applied Physiology} 86, 359 (1999).

\bibitem[14]{Tisean} R. Hegger, H. Kantz, and T. Schreiber, Practical
implementation of nonlinear time series methods: The TISEAN package\textit{,}
\textit{CHAOS} 9, 413 (1999).

\bibitem[15]{Kennel} M. B. Kennel, R. Brown, and H. D. I. Abarbanel,
Determining embedding dimension for phase-space reconstruction using a
geometrical construction\textit{,} \textit{Phys. Rev. A} 45, 3403 (1992).

\bibitem[16]{Farmer} J. D. Farmer, and J. J. Sidorowich , Predicting chaotic
time series. \textit{Phys. Rev. Lett}. 59, 845 (1987).
\end{thebibliography}
\end{document}